\documentclass[letterpaper,twocolumn,10pt]{article}
\usepackage{usenix-2020-09}

\usepackage{tikz}
\usepackage{amsmath}

\usepackage{algorithm}
\usepackage{algpseudocode}
\usepackage{booktabs}
\usepackage{makecell}
\usepackage{multirow}
\usepackage{caption}
\usepackage{enumitem}
\usepackage{endnotes}

\newcommand{\SYS}{FusionStitching}
\newcommand{\NAME}{\textit{\SYS} }

\newcommand{\parahead}[1]{\textbf{\textit{#1}}}

\begin{document}

\date{}

\title{FusionStitching: Boosting Memory Intensive Computations for \\ Deep Learning Workloads}

\author{
{\rm Zhen Zheng, Pengzhan Zhao, Guoping Long, Feiwen Zhu, Kai Zhu,} \\ 
{\rm Wenyi Zhao, Lansong Diao, Jun Yang, Wei Lin} \\
{\normalsize Alibaba Group} \\
{\normalsize \{james.zz, pengzhan.zpz, guopinglong.lgp, feiwen.zfw, tashuang.zk,} \\
{\normalsize kevin.zwy, lansong.dls, muzhuo.yj, weilin.lw\}@alibaba-inc.com}
}


\maketitle

\setlist[itemize]{noitemsep, topsep=0pt} 
\setlength{\belowdisplayskip}{0pt}
\setlength{\belowdisplayshortskip}{0pt}
\setlength{\abovedisplayskip}{0pt}
\setlength{\abovedisplayshortskip}{0pt}

\begin{abstract}

We show in this work that memory intensive computations can result in severe 
performance problems due to off-chip memory access and CPU-GPU context switch 
overheads in a wide range of deep learning models. 
For this problem, current just-in-time (JIT) kernel fusion and code generation techniques 
have limitations, such as rough fusion plan exploration strategies and limited code generation ability. 
We propose \NAME, a deep learning compiler capable of fusing memory intensive 
operators, with varied data dependencies and non-homogeneous parallelism, 
into large GPU kernels to reduce global memory access and context switch overhead automatically. 
\NAME widens the range of operation combinations that fusion can target beyond previous JIT works 
by introducing data reuse of intermediate values.
It explores large fusion spaces to decide optimal fusion plans 
with considerations of memory access costs, kernel calls and resource usage constraints.
\NAME tunes the optimal stitching scheme with a domain-specific cost model efficiently.
Experimental results show that \emph{\SYS} can reach up to $2.21\times$ speedup compared to state-of-the-art, with 1.45$\times$ on average.
Besides these experimental results, we integrated our approach into a compiler 
product and deployed it onto a production cluster for AI workloads with thousands of GPUs.
The system has been in operation for more than 4 months 
and saves 7,000 GPU hours on average for approximately 30,000 tasks per month.


\end{abstract}



\section{Introduction}
\vspace{-1ex}

Recent years have witnessed a surge of industry scale applications of deep learning models,
ranging from images/videos, text/NLP, to billion scale search and
recommendation systems\cite{jizhe}. Such workloads are typically expressed as computation 
graphs, and mapped to hardware 
through domain specific frameworks (TensorFlow\cite{tensorflow}, 
PyTorch\cite{torch}, MXNet\cite{mxnet}, etc). Given the flexibility and expressiveness 
of modern execution frameworks, there are still challenges regarding to transforming 
high level computation graphs into efficient kernels to maximize the underlying hardware 
execution efficiency.

Many current research works mainly focus on \textbf{compute intensive} operations, referring to 
GEMM and convolution in this paper, as compute intensive operations dominate the 
execution time for many DNN workloads \cite{halide,tc,tvm,astra,pbqp}, such as CNN\cite{vgg,resnet,inception}). 
Compute intensive operations are usually realized as efficient libraries (cuDNN, cuBLAS).

However, recent advancement of deep learning domain has resulted in many novel 
model structures in which memory intensive patterns occupies a large proportion of 
time. In this paper, we refer to operators that are not GEMM or convolution as 
\textbf{memory intensive} ops, such as element wise\cite{tfmathmuldoc}, 
transpose\cite{xlatransposedoc} and reduction\cite{xlareducedoc}. 
In addition, the amount of memory intensive operators in modern machine learning 
models can be very large, causing notable GPU kernel launch and framework scheduling 
overhead. Table \ref{tbl:kernel-breakdown} contains the collected metrics of various 
models with TensorFlow implementation. The execution time of memory intensive 
ops can be more than that of compute intensive ops in some cases, 
and the kernel calls can be up to 10,406.
Optimizing compute intensive ops alone 
is inadequate to unlock the full performance potential for these models.

It is not feasible to build library for memory intensive operations,
because a single memory intensive op is too simple while the combination of 
such ops various in different models and changes fast as the model evolves.
Thus, memory intensive ops are usually generated just-in-time with compilation 
techniques in modern machine learning frameworks.

A common approach to address memory intensive patterns is computation fusion. 
Prior works have explored the basic idea in AI workloads\cite{ppopp2015, ma2020rammer},
database\cite{kernelweaver}, image processing\cite{cgo2019, halide, video}, and HPC 
applications\cite{hpcfusion, ode-fusion} ahead-of-time (AOT). 
However, how to fuse kernels just-in-time (JIT) efficiently, with unpredictable varied dependencies and non-homogeneous 
parallelism, is still an open problem.
Note that the rapidly evolving AI models introduce 
diverse and complex combination patterns of ops. 

Existing JIT kernel fusion techniques use simple code generation and fusion exploration approach.
As for XLA\cite{xla}, state-of-the-art JIT optimization engine, 
it only focuses on memory access optimization for memory intensive ops, 
but lacks consideration of computation characteristics.
In a fusion kernel generated by XLA, each thread is only capable of reading 
intermediate results produced by itself. 
If two threads require the same intermediate value, they will recompute it independently.
It works good for light element-wise ops (like add, sub, mul), but introduces severer 
re-computation overhead for ops like reduction, tan, log and other expensive ops.
XLA avoids re-computation overhead by only allowing expensive ops (reduction, tan, et.al.) 
appear in the tail of a fusion, that is not being a producer within a fusion.
This trade-off limits the fusion exploration space.
Meanwhile, XLA uses a conservative fusion exploration strategy. 
On one hand, it is a rule-based strategy which cannot cover the various combination 
of element wise ops and tensor shapes.
On the other hand, it uses a greedy approach that is easily to fall into local solution.

We propose \NAME, a JIT optimization framework to systematically 
perform fusion space exploration and generate high-performance kernels efficiently.
\NAME broadens the fusion exploration space by exploring data reuse between ops.
If the intermediate data can be reused by a set of threads, 
these threads could exchange the intermediate data through register-shuffle and shared memory.
This approach allows expensive ops to be fused in the middle of a kernel and widens the fusion possibility.
With the expanded fusion exploration space, \NAME is able to compose a large set of 
ops with diverge and complex patterns into one GPU kernel.
This is effective to reduce off-chip memory accesses and context switch overhead. 
Meanwhile, \NAME applies a more effective fusion exploration approach to find good fusion patterns.
\NAME addresses two main challenges of large scope JIT fusion.


The first challenge is how to generate efficient GPU kernel given 
unpredictable complex \textit{fusion pattern} just-in-time. 
A \textbf{fusion pattern} reveals a set of operators to be fused into a single GPU kernel.
It is non trivial to handle a complex fusion pattern consisting of a broad range of 
memory intensive ops with various dependence relationships and tensor dimensions. 
We provide 4 types of stitching scheme abstractions covering the main 
patterns for memory intensive ops in machine learning workloads, 
including independent packing, re-computation, intra-warp reuse and intra-block reuse scenarios. 
With the stitching schemes, we design an approach that divides a fusion into several sub-groups.
Different sub-groups use independent schedules and 
adjacent sub-groups communicate with proper stitching scheme.
\NAME explores sub-group and stitching scheme settings automatically for a fusion pattern, 
along with launch dimension enumeration. 
Instead of using rule-based approach, we design a cost model for code generation tuning.

The second challenge is to find the optimal \emph{fusion plan} given complex op graph.
A \textbf{fusion plan} reveals how ops in a subgraph are grouped together 
to form a set of \textit{fusion patterns}.
Note that naive composition of multiple 
computations may cause notable performance slowdown, as different portions of 
the kernel may have conflicting memory layout, parallelization and on-chip resource
requirements\cite{versapipe}. 
A deep learning computation graph usually brings 
huge search space about operation combinations. It is not feasible to evaluate 
all possible combinations with complexity of up to $O(2^{V})$ where $V$ is 
the number of ops in the computing graph. We formulate the fusion plan searching as an approximate dynamic 
programming problem with complexity of $O(V+E)$, where $E$ is the number of edges 
in the graph. The approximate dynamic programming process produces a limited set of 
promising fusion patterns with a light-weight cost model.
\NAME applies beam search to generate 
the overall fusion plan with these fusion patterns.


We realize \NAME into TensorFlow as an optimization component beyond XLA.
All the optimizations are opaque to users. \NAME supports JIT optimization for both training 
and inference.

We evaluate \NAME on a set of common machine learning models, ranging from  
natural language processing, OCR, speech recognition to searching and recommendation models. 
\NAME achieves up to 2.21$\times$ speedup compared with XLA, with 1.45$\times$ on average.

In summary, this work makes the following contributions:

\begin{itemize}

\item It reveals the importance of memory intensive ops, 
and broadens the range of op combinations that JIT fusion targets by 
introducing data reuse of to-be-fused ops.

\item It provides a set of fusion scheme abstractions for memory intensive ops in machine 
learning workloads, 
and proposes an approach to generate efficient GPU kernels 
given very complex fusion patterns.

\item It proposes a technique to explore good fusion plans just-in-time in 
the expanded search space of op composition, 
along with a two-layer cost model of GPU kernels.

\item It provides an industry level realization that is opaque to users and evaluates with 
various modern machine learning models on production cluster.

\end{itemize}

\vspace{-1ex}
\section{Motivation and Challenge}
\label{section:motivation}
\vspace{-1ex}

\subsection{Motivation}
\label{subsec:jit-fusion}
\vspace{-.5ex}

JIT fusion is the state-of-the-art technique to reduce context switch and memory access overhead of intensive memory operations for machine learning workloads. 
As the state-of-the-art fusion engine, XLA only supports thread-local data transferring for fusion, which relies on index analyzing and re-computation to improve thread locality.
A bad case is to put a reduction in the middle of a fusion pattern.
If different threads require the same value of reduction result, each thread needs to recompute the reduction independently and redundantly.
To avoid this, XLA does not fuse operations that lead to middle-reductions. As a result, this strategy misses many optimization exploration space.

\begin{figure}[h!]
    \centering
    \includegraphics[width=\columnwidth]{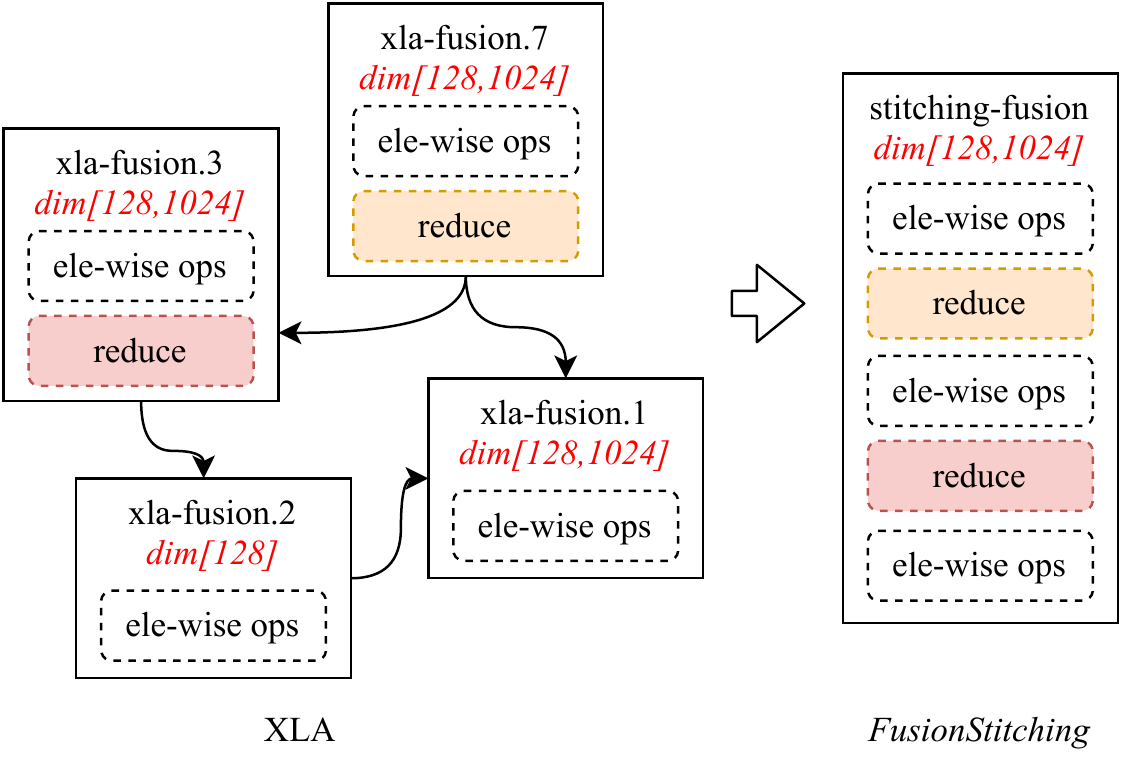}
    \vspace{-3ex}
    \caption{The difference of fusion pattern between XLA and \NAME for \textit{Layer Normalization}.}
    \vspace{-1.5ex}
    \label{fig:fusion-pattern-case}
\end{figure}

Figure~\ref{fig:fusion-pattern-case} shows how XLA and \NAME perform fusion for layer normalization. XLA forms 4 fusions, where \textit{fusion.3} and \textit{fusion.7} 
end with reductions and \textit{fusion.2} ends with expensive operation.
XLA does not do other fusion to avoid re-computation overhead, 
with the drawback of CPU-GPU context switch and off-chip memory access overhead. While \NAME generates only one kernel with all intermediate results residing in on-chip memory and involves no CPU-GPU switching. No re-computation is introduced in \NAME.

Static libraries are not feasible to solve this problem as a single memory intensive operation is lightweight and the combinations of them vary in different models.
TVM~\cite{tvm,zheng2020ansor} mainly focuses on compute intensive operations but not fusion optimization, 
whereas XLA only provides simple rule-based fusion strategies for memory intensive ops. 
In addition, TVM relies on pre-defined schedules for code generation and usually requires a long time for tuning.

\subsection{Observation}
\label{subsec:motivation}
\vspace{-.5ex}

In this work, we have analyzed a variety of machine learning workloads (Section~\ref{section:evaluation}).
we get two main observations.

\parahead{Severe Context Switch Overhead}
Context switch overhead between CPU and GPU is a severe performance issue. 
Table~\ref{tbl:kernel-breakdown} shows the breakdown analysis of a wide range of 
machine learning applications.
The count of GPU kernel calls ($\#$) for TensorFlow 
implementation can reach up to 10,406, which leads to large kernel launch overhead. 
Even fused with XLA, the kernel calls can be still up to 6,842. 
As a result, both scheduling and pre-/post-processing time on CPU brought by machine learning framework dominate the execution time for some models (like \textit{BERT} inference, \textit{DIEN}, \textit{ASR}, and \textit{CRNN}).

\parahead{Large Portion of Memory-intensive Ops}
Also in Table~\ref{tbl:kernel-breakdown}, the  
execution time of memory intensive ops can be up to 40\% in the overall time for some models. Off-chip memory traffic time usually occupies a large portion of the overall execution time of memory intensive ops. By fusing a large quantity of operations, our system is able to reduce CPU-GPU context switch overhead and leverage high speed on-chip memory to reuse intermediate data between ops.

\subsection{Challenges}
\vspace{-.5ex}

Although data reuse is a potential opportunity to fuse a large scope of ops together to avoid redundant re-computation, there are several main challenges for both applying reuse given a fusion pattern and deciding op fusion strategy.

As for applying reuse for a given fusion pattern to generate kernel:

\begin{itemize}

    \item The first challenge is how to evaluate reuse benefit. Reuse is not always better than re-computation. Reuse will lead to extra inter-thread data communication, whereas re-computation introduces redundant compute overhead but without inter-thread communication. Besides, introducing reuse with shared memory may hurt kernel occupancy, thus hurts parallelism.

    \item The second challenge is how to apply reuse. There are two types of data reuse: intra-warp reuse and intra-block reuse. \textbf{Intra-warp/block} reuse means that threads within a warp/block reuses intermediate data  produced by other threads within the same warp/block. The trade-off is that, intra-warp reuse has less memory access overhead while intra-block reuse has better parallelism.

\end{itemize}

For a given machine learning model, the challenge is to decide what ops should be fused together. XLA also tries to solve this challenge, but with limited exploration space. One main problem is that, forming a fusion pattern by data reuse is not always better than separate kernels.
Reuse requires data locality within thread-block or warp, which can potentially limit parallelism.
Take reduction as an example, reuse of reduction result requires to compute reduction within a block or warp.
However, some reductions perform better with several blocks work together, which has higher parallelism.
Not fusing such a reduction with its consumers may result in better performance, even with context switch and off-chip memory access overhead. 
Intra-block reuse may further hurt parallelism as it requires extra shared memory.
Another typical problem is that, when operation A can be fused with either operation B or operation C, while B and C cannot be fused together due to some limitation, 
it is not always easy to decide which one to fuse for A. 
Rule-based approaches, like XLA, fail to find effective fusion plans for varied models.

\vspace{-.5ex}
\section{Overview}
\label{section:overview}
\vspace{-1ex}

\subsection{Data Reuse}
\label{subsec:data-reuse}
\vspace{-.5ex}

As described before, \NAME widens fusion possibilities by introducing data reuse,
a rarely used method in state-of-the-art JIT fusion techniques.
Data reuse can be applied when different threads processing operation $B$ 
requires the same value generated by operation $A$. 
We assume $B$ is the consumer of $A$.
The threads requiring the same value can read the same data with inter-thread communication,
but do not have to re-compute it redundantly like in XLA.
We observe that tensor shapes of a sequence of memory intensive operations 
always shrink and broaden frequently due to operations like \textit{reduce}, 
\textit{broadcast}, \textit{gather}, and \textit{slice}.
Thereby the opportunity to explore data reuse is large.

In our work, we explore both intra-warp reuse and intra-block reuse, 
corresponding to \textit{warp composition} and \textit{block composition} 
in figure~\ref{fig:composition-schemes}. 
Take reduction as an example, intra-warp reuse means that each warp does reduction 
for a row of data and stores result in the register of the first lane of the warp. 
Consumers of the reduction read data with register-shuffle from the first lane.
Intra-block reuse does reduction for the row with all threads in the block 
and stores results in shared memory.
Consumers of the reduction read data from shared memory. 
This approach enables stitching operations with on-chip memory while avoiding re-computation. 
Data locality should be guaranteed for correct data reuse. 
Intra-warp reuse requires warp level data locality, 
and intra-block reuse requires block level data locality.

\begin{figure}[h!]
    \centering
    \includegraphics[width=\columnwidth]{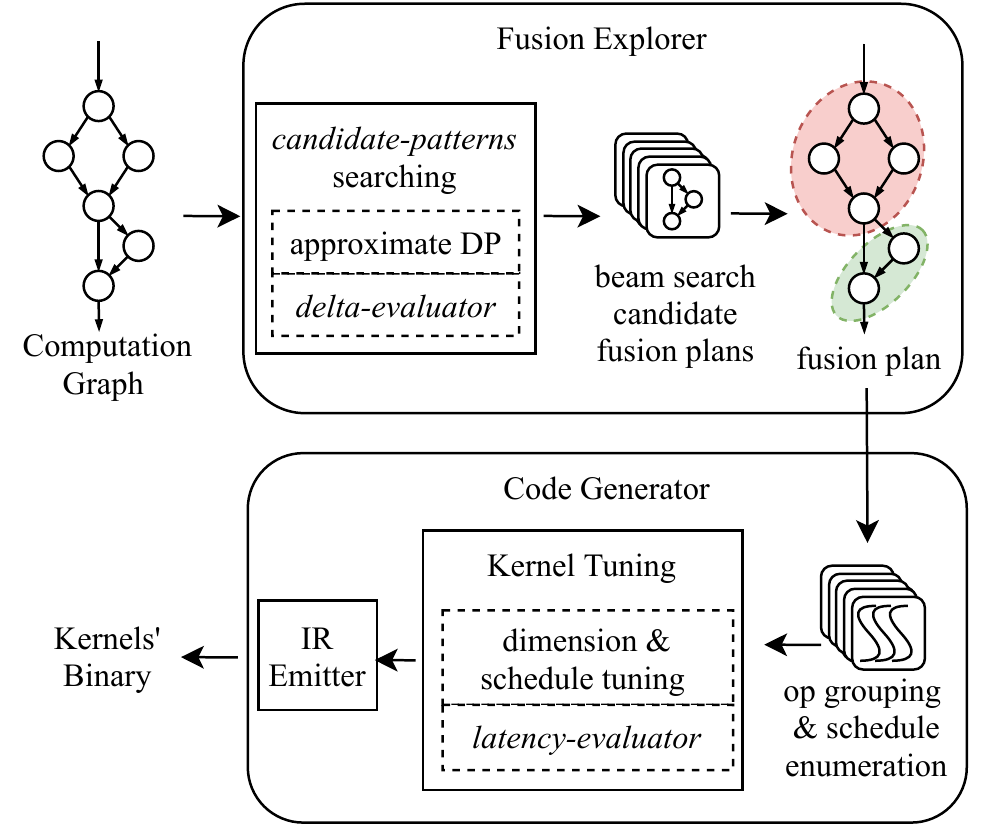}
    \vspace{-3ex}
    \caption{Overview of \NAME.}
    \vspace{-2ex}
    \label{fig:overview}
\end{figure}

\subsection{\NAME System}
\vspace{-.5ex}

We divide optimization process into two stages in \NAME:
fusion exploration and code generation. 
The two stages are conceptually independent but highly related.

As shown in figure \ref{fig:overview}, \NAME consists of \textit{fusion explorer}
and \textit{code generator}. 
\textit{Fusion explorer} generates possible fusion patterns that may enjoy data reuse. It then applies beam search to form several candidate fusion plans with these fusion patterns, 
and selects the best fusion plan from candidate plans with a cost model.
\textit{Code generator} generates GPU kernels for each fusion pattern 
produced by \textit{fusion explorer}. 
We pre-define a set of schedules (i.e., templates) for each kind of operators. 
A schedule describes code generation about how an operation runs in parallel. 
Meanwhile, it indicates whether threads of an operation reuse data with other threads and the reuse scope, 
and whether this operation produces data for consumer's reuse.
\NAME divides operations of a fusion pattern into several groups, 
and enumerates all valid combinations of the schedules in the groups.
With a cost model that estimates the performance of each schedule and launch dimension setting, 
\NAME selects the best configuration of the fusion pattern and generates GPU kernel code.

We design a two-level cost model in \NAME. \textit{Fusion explorer} needs to 
search in large search space and applies \textit{delta-evaluator} (\ref{subsec:fusion-eval}), which is fast 
but less accurate. 
\textit{Code generator} operates on merged GPU kernels and needs more accurate performance estimation, 
and thus we apply \textit{latency-evaluator} (\ref{subsec:cost-model-I}) which is more accurate but slower.

We first describe how \textit{code generator} works in section~\ref{section:codegen} and then describe \textit{fusion explorer} in section~\ref{section:fusion}.
\vspace{-.5ex}
\section{Code Generation}
\label{section:codegen}
\vspace{-1ex}

\textit{Code generator} takes a fusion pattern as input, and produces a GPU kernel 
that implements the fused operators. 
It is nontrivial to fuse multiple ops into one 
high performance GPU kernel due to various dependence scenarios and parallelism incompatibilities.

The combination patterns of memory intensive operators in machine workload are numerous, 
but basic kinds of memory intensive ops are limited.
We made three classifications: light element-wise (most elem-wise ops), 
expensive element-wise (tan, exponential, et.al.), and reduction ops.
We pre-define a set of \textit{schedules} for each kind of memory intensive ops. 
The left problem of code generation 
is how to stitch different operators into one kernel and what schedule each individual 
op applies.

We first systematically investigate four kernel composition schemes (\ref{subsec:codegen-schemes}) 
that covers common execution patterns for memory intensive operations.
With these composition schemes, 
we used an automatic generation solution based on performance modeling (\ref{subsec:cost-model-I})
to find good schedules and generate code for the fusion pattern (\ref{subsec:codegen-opt}).



\subsection{Kernel Composition Schemes}
\label{subsec:codegen-schemes}
\vspace{-.5ex}

\begin{figure}
\centering
\includegraphics[width=\columnwidth]{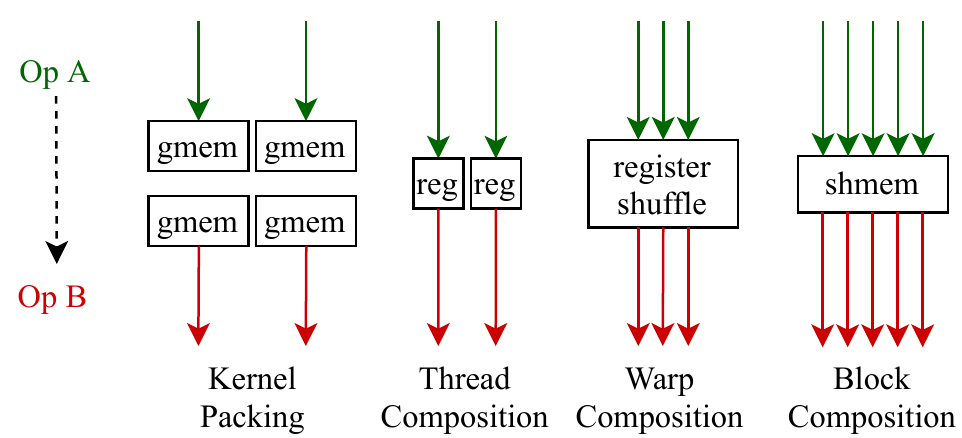}
\vspace{-3ex}
\caption{The four types of composition schemes when stitching op A and B together.
\textit{Gmem}: global memory (different square indicates different memory address).
\textit{Reg}: register. \textit{Shmem}: shared memory.}
\vspace{-2ex}
\label{fig:composition-schemes}
\end{figure}

We study about four kernel composition schemes, which indicate main behaviors of common 
memory intensive ops. Different scheme indicates different data dependence 
and parallel behaviors of kernels to fuse, ranging from no dependence to complex 
cross-thread dependence, and from uniformed parallelism to non-homogeneous computations. 
Figure \ref{fig:composition-schemes} illustrates the four kind of composition schemes.

\parahead{Kernel Packing} packs computations of ops with no data dependence.
This scheme is instrumental in reducing context switch overhead of kernel launch and 
framework scheduling. It also reduces loop control overheads in instruction level.
To reduce control flow overhead, we perform aggressive loop fusion \cite{Kennedy,gao} 
to merge as many element-wise ops as possible into a single loop structure when 
these ops have the same parallelism dimension.

\parahead{Thread Composition} fuses dependent ops and 
transfer intermediate results via registers within a local thread context. 
This is the fusion scheme XLA applies. 
This may introduce redundant computations between threads requiring the same value.

\parahead{Warp Composition} fuses dependent operators and apply intra-warp reuse. 
This scheme employs register shuffle to communicate between threads within 
a warp. A typical case is warp reduction and its consumers, 
which applies warp level reduction and leverages registers to transfer 
intermediate results to threads of its consumers.

\parahead{Block Composition} applies intra-block reuse and unlocks the potential to 
enable composing non-homogeneous computations into large fused kernels, 
as long as these computations can communicate within block level. 
It makes use of shared memory to transfer intermediate results for data reuse. 

\textit{Warp composition} and \textit{block composition} are flexible schemes 
as they allow different ops to execute in independent schedules in the fused kernel. 
The only requirement is to keep warp/block locality between producers and consumers.
These two schemes are essential to compose a broad range of op kinds with various 
parallelism characteristics and dependence relationships efficiently.

As far as we know, \NAME is the first work to thoroughly study about all 
above composition techniques for just-in-time compilation of memory intensive 
operations for machine learning workloads.
XLA only realizes kernel packing and thread composition. 
We do not stitch ops that involves inter-block communications as it results in 
global memory level synchronization and introduces high overhead.

\subsection{Kernel Generation}
\label{subsec:codegen-opt}
\vspace{-.5ex}

Code generation in \NAME is not as straightforward as XLA, 
because of the trade-off of composition schemes we discussed above.
We use a cost model based tuning approach in \NAME, and ease the process by op grouping.

According to the four types of composition schemes, 
we pre-define a set of schedules for each kind of operators (light element-wise, 
expensive element-wise, and reduction ops).
Specifically, we define a single template for light element-wise ops representing 
\textit{kernel packing} and \textit{thread composition}.
Note that we find \textit{kernel packing} and \textit{thread composition} 
can be described with the same schedule.
We define three schedules for expensive element-wise ops and reduction ops.
The first one represents \textit{kernel packing} and \textit{thread composition}.
The second one stores intermediate result to the register of the first lane in each warp,
representing \textit{warp composition}.
The third one stores intermediate result to shared memory, representing \textit{block composition}.

\NAME enumerates all schedule combinations of ops. 
It estimates these combinations according to cost model (sec~\ref{subsec:cost-model-I}). 
To ease the enumeration process, we divide ops in a fusion pattern into several groups. 
Each group applies one schedule and different groups transfer 
intermediate data with warp/block level reuse.

We group ops according to op types.
Given a fusion pattern, we enumerate a small set of grouping strategies.
We call \textit{sub-root} as the output op of a group, and \textit{root} as the output of the fusion.
We first identify sub-roots and generate groups rooted from sub-roots and root.
Reduce ops are always regarded as sub-root. 
Expensive element-wise ops are enumerated to both sub-roots and non sub-roots.
Other ops are neither sub-roots.
This leads to a set of groups rooted from reduction, expensive element-wise and root ops.

The insight of the grouping approach is that, the schedule of non sub-roots 
can be determined by the schedule of sub-roots by tensor indices propagation. 
Thus we only need to enumerate the schedule of sub-roots and root ops, 
and we can propagate to get schedules of all ops.

\NAME enumerates grouping strategies, and emulates schedules of every sub-root/root op
and launch dimension of the fused kernel.
As data reuse requires correct data locality in the reuse scope, 
schedules do not match data locality requirement are discarded.
After estimating the performance of each enumeration with \textit{latency-evaluator}, 
\NAME selects code generation strategy with the best estimated performance.

\vspace{-.5ex}
\subsection{Kernel Evaluation: Latency-Evaluator}
\label{subsec:cost-model-I}
\vspace{-.5ex}

\NAME requires an accurate estimation for kernel performance for code generation strategy.
Fortunately, as the searching space of code generation is not very large, 
we can tolerate a relative slow cost model.

We build \textit{latency-evaluator} as in equation~\ref{eq:cost-model-I}, where 
$L$ is the estimated execution cycles of a fused kernel.

\begin{equation}
\begin{aligned}
\label{eq:cost-model-I}
L=N_{wave} \times L_{warp} \\
N_{wave}=\frac{N_{warp}}{Occupancy} \\
L_{warp}=N_{instruction} \times CPI
\end{aligned}
\end{equation}

$N_{wave}$ means how many waves of warps that will be processed by a GPU, 
noting that the warps for a large GPU kernel will be spitted into several 
waves to be executed on a GPU where warps in the same wave executes 
concurrently. $L_{warp}$ means the latency (cycles) a single warp spends 
in the fused kernel. The multiply of $N_{wave}$ and $L_{warp}$ stands for 
the total cycles to execute the fused kernel.

We estimate $N_{wave}$ with the total number of warps to issue (${N_{warp}}$) 
and the occupancy of the fused kernel (${Occupancy}$). $N_{warp}$ is calculated 
with launch dimension and tensor shape. We calculate $Occupancy$ with launch dimension, shared 
memory usage (Sec.~\ref{subsec:shared-memory-optimization}) and estimated register usage.
We estimate the register usage by analyzing the life time of every intermediate 
value and get the maximum register usage for a thread. 
Also, we estimate shared memory usage by life time analyzing.
This approach is accurate enough for us to calculate $Occupancy$.

As for $L_{warp}$, we use the reported $CPI$ numbers for different types of ops
\cite{jia2018dissecting,jia2019dissecting} 
and multiply it with the total instruction count ($N_{instruction}$) we estimated.

\vspace{-.5ex}
\subsection{Shared Memory Optimization}
\label{subsec:shared-memory-optimization}
\vspace{-.5ex}

It is essential to use shared memory moderately as large amount of shared memory usage 
hurts kernel parallelism, especially for large granularity compositions.
To use as much shared memory as possible while not hurting parallelism, we explore 
a dataflow based shared memory sharing technique. The insight is that, 
\NAME reuses previous allocated shared memory as much as possible to reduce 
unnecessary shared memory allocation.

We use dominance tree algorithm\cite{dominance-tree} for shared memory dataflow 
analysis. The approach takes a computation graph and shared memory requests as input, 
and outputs an allocation map. To optimize shared space sharing, we traverse ops of 
the computation graph in topological order. When an op does not need shared space, 
previous allocation information will be propagated forward. If an op needs shared 
space, we merge allocation information of all its operands, test the dominance 
relation to check if we can share any previously allocated space for current op, 
and reuse the space if possible.

\vspace{-.5ex}
\subsection{Computation Reuse Optimizations}
\vspace{-.5ex}

Beside the reuse of intermediate result between producers and consumers
with shared memory and register shuffles,
\NAME also reduces thread local redundant calculations.
Thread local redundant calculations mainly come from 
memory access index calculations and thread local intermediate values.
It is because different parts in a fusion kernel may use different schedules, 
and index and some intermediate values are generated independently within each schedule, 
even in the same thread.
Before generating the code, \NAME first analyzes the overall index and intermediate 
value characteristics and then organizes the output code to reuse computations and 
data as much as possible.

\vspace{-.5ex}
\section{Fusion Exploration}
\label{section:fusion}
\vspace{-1ex}

As noted before, rule-based approach in XLA is not adequate to handle complex fusion decisions. 
We use a cost-based searching approach to find good fusion plans.
We propose an approximate dynamic programming approach to search for a set of promising 
fusion patterns(\ref{subsec:fusion-explore}) which may overlap with each other. 
The searching process is guided by a light-weight domain-specific cost model(\ref{subsec:fusion-eval}). 
\NAME finally generates the overall fusion plan by selecting fusion patterns(\ref{subsec:fusion-plan}).

\subsection{Fusion Problem Definition}
\label{subsec:fusion-problem}
\vspace{-.5ex}

We formulate fusion exploration as a subgraph searching problem. 
For computation graph $G=(V,E)$, where $V$ and $E$ are sets of vertices and edges respectively. 
We define a fusion pattern $P_i=(V_i,E_i)$ as a subgraph of $G$, with $V_i \subseteq V$, $E_i \subseteq E$. 
A fusion plan is a set of disjoint fusion patterns $S=\{P_0,\cdots,P_{k-1}\}$. 
The score function of $P_i$ is annotated as $f(P_i)$. The higher the performance is, the larger $f(P_i)$ is.
So the goal of computation fusion problem is to find fusion plan $S$ with maximal $\sum_{i=1}^{k} f(P_i)$.

\subsection{Explore Fusion Patterns}
\label{subsec:fusion-explore}
\vspace{-.5ex}

A Brute-force way to enumerate all fusion patterns has a complexity of up to $O(2^{V})$ without pruning. 
We proposed an \textit{approximate} dynamic programming approach with complexity 
of $O(V+E)$ to find good fusion patterns.

The basic idea of fusion exploration is that, we generate 
\textit{candidate-patterns} starting from each vertex in post-order in the graph, 
and select and compose final fusion plan with these candidate patterns. 
The \textbf{\textit{candidate-patterns}} starting from vertex $V_i$ is the set of  
fusion patterns whose producer node is $V_i$. 
We describe how we generate \textit{candidate-patterns}  
for each vertex in this section and describe how to compose the final fusion plan in 
section \ref{subsec:fusion-plan}

Given a computation graph $G$, we get a topological sorting list. 
We generate \textit{candidate-patterns} for vertices in post-order, from the last vertex 
to the first vertex. 
This is to guarantee that each searching results in a fusion pattern 
with current vertex as producer.

Equation~\ref{eq:fusion-exploration} shows the \textit{approximate} dynamic programming process.
$C_i$ is the set of all consumers of $V_i$. 
\textbf{PatternReduction} is the approach to get candidate patterns of $V_i$ 
according to all its consumers' candidate pattens.

\begin{equation}
\begin{aligned}
\label{eq:fusion-exploration}
\vspace{-1ex}
P_i=PatternReduction(C_i)
\vspace{-1ex}
\end{aligned}
\end{equation}

\begin{figure}
    \centering
    \includegraphics[width=\columnwidth]{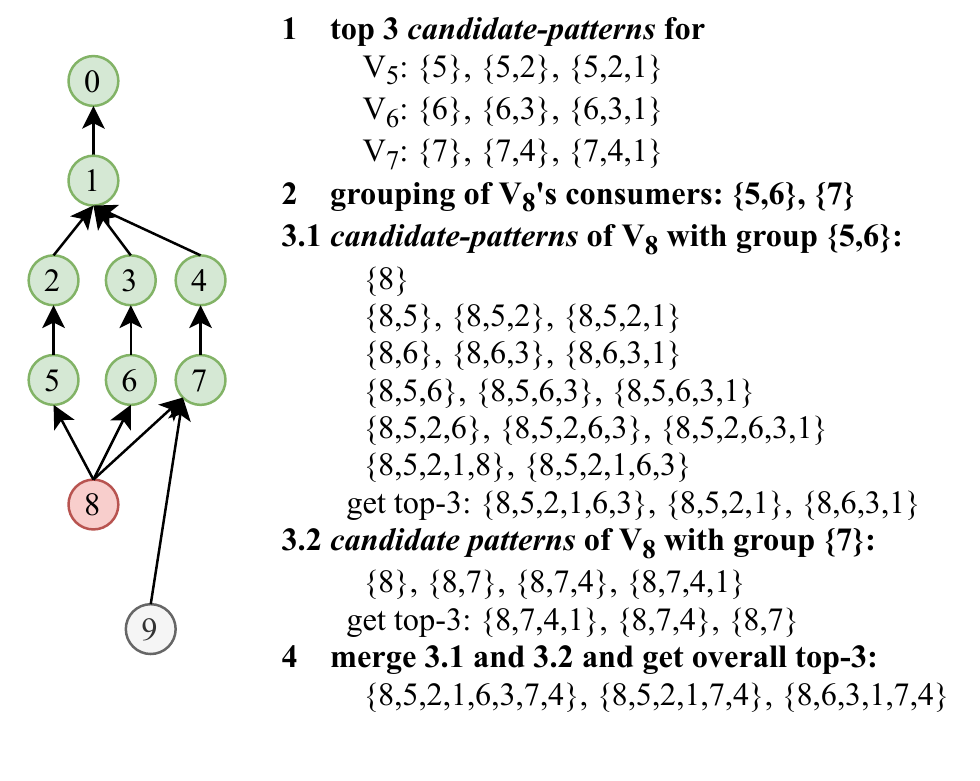}
    \vspace{-4ex}
    \caption{\textit{PatternReduction} case to generate \textit{candidate-patterns} for $V_8$.}
    \vspace{-2.5ex}
    \label{fig:group-reduction-case}
\end{figure}


We describe \textit{PatternReduction} approach with an example shown in Figure \ref{fig:group-reduction-case}, 
where $V_0$ is the \textit{root} vertex who produces the output of whole graph.
Assume that \textit{candidate-patterns} for vertices before $V_8$ have already been generated.
(We only explore top $k$ patterns in \textit{candidate-patterns} for 
each vertex according to score function $f$. 
We set $k$ as 3 in this case.)
We will generate \textit{candidate-patterns} for $V_8$ with its consumers' information. 
A naive approach is to append $V_8$ to all possible combinations 
of patterns in its consumers' \textit{candidate-patterns} sets, and select the top 3 
patterns within the appended combinations. 
However, the combinations will be huge when the consumer number and top $k$ setting is large, 
noting that JIT approach requires timely optimization.
Instead, we design an \textit{approximate divide-and-conquer} process to 
find top 3 patterns with limited complexity as \textit{PatternReduction}.

We first divide the consumers of $V_8$ into several groups and find \textit{candidate-patterns} 
for $V_8$ considering these groups independently, 
and finally compose final \textit{candidate-patterns} by reducing the above results of all groups. 
We assume the group number in this case is 2 (group
\{$V_5$, $V_6$\} and group \{$V_7$\}). For group \{$V_5, V_6$\}, we enumerate all 
possible combinations of patterns in \textit{candidate-patterns} of $V_5$ and $V_6$. 
Specifically, there are 15 possible combinations between $V_5$ and $V_6$, including empty set. 
We append $V_8$ to each of the 15 combinations and select the top 3 patterns 
as the temporary candidates associated to group \{$V_5, V_6$\}. 
We get another top 3 patterns considering group \{$7$\} as temporary candidates.
We finally select the final top 3 patterns as \textit{candidate-patterns} for $V_8$
from all above 6 temporary candidates. 
Note we validate top patterns according to score function $f$.

The \textit{PatternReduction} process above is recursive 
if consumers number of a vertex is very large. 
We first divide the consumers into two groups.
To get \textit{candidate patterns of a group}, we divide the group further until it is small enough.

\begin{figure}
    \centering
    \includegraphics[width=\columnwidth]{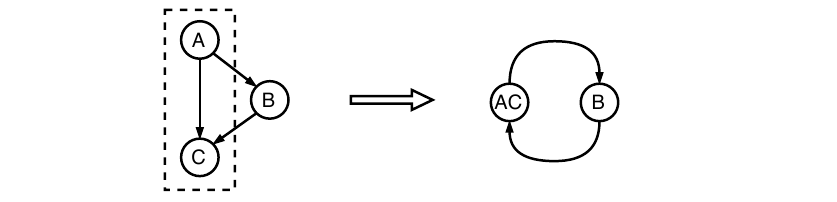}
    \vspace{-3ex}
    \caption{Cyclic Dependence: a cyclic dependence occurs after fusing nodes $A$ and $C$ together}
    \vspace{-3ex}
    \label{fig:cycle}
\end{figure}

A constraint of a fusion pattern is that, no cyclic dependence is allowed.
Figure \ref{fig:cycle} shows an example that a cyclic dependence occurs after fusion. 
\NAME discards patterns with cyclic during the searching process.
Meanwhile, \NAME only explores fusion patterns that the \textit{code generator} can 
process. It does not form fusion patterns with cross-block communication requirement.

\begin{figure}
    \centering
    \includegraphics[scale=0.8]{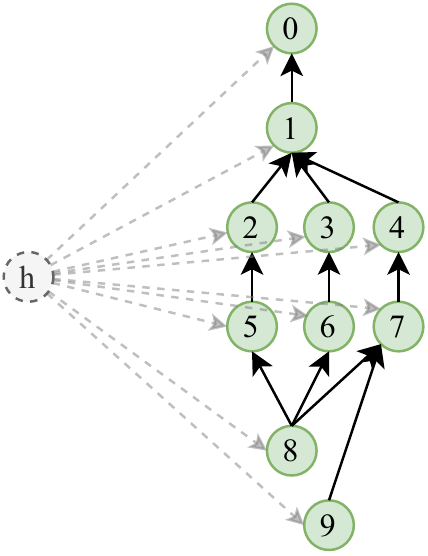}
    \vspace{-2ex}
    \caption{Fusion merge of remote patterns.}
    \vspace{-1ex}
    \label{fig:horizontal-fusion}
\end{figure}

\parahead{Remote Fusion}
To further reduce the context switch overhead between CPU and GPU, 
we try to merge fusion patterns that are not adjacent in the graph after above procedures.
As is shown in Figure \ref{fig:horizontal-fusion}, we add a virtual vertex $h$ 
as the producer for all vertices and apply \textit{PatternReduction}. 
We finally get the \textit{candidate-patterns} of $V_h$, 
which includes the fusion of remote patterns that are not adjacent.
The remote pattern fusion helps to reduce generated kernels and 
thus reduces the context switch overhead.
\textit{Remote fusion} results in \textit{kernel packing} for code generation.

\subsection{Generate Overall Fusion Plan}
\label{subsec:fusion-plan}
\vspace{-.5ex}

All \textit{candidate-patterns} of all vertices, which may overlap with each other, form a new set $E$. 
The insight of forming good fusion plan is to select non-overlapped patterns 
from $E$ with high score of $f$. 
Note $f$ means how much cost the pattern saves (sec~\ref{subsec:fusion-eval}).

\NAME uses beam search~\cite{furcy2005limited} to generate top-3 (width of the beam) candidate fusion plans,
and finally selects the best plan within the 3 candidates 
with \textit{latency-evaluator} (sec~\ref{subsec:cost-model-I}).

Specifically, \NAME maintains 3 \textit{buffer sets} to store candidate fusion plans.
It traverses from the producer vertex to the consumer vertex in op graph $G$, 
and try to append each \textit{candidate pattern} of each vertex to each \textit{buffer set} 
in turn if it introduces no overlapping.
It may result in more than 3 temporal sets after appending in each iteration, 
and will select top-3 with highest accumulated $f$ as new \textit{buffer sets}.

After traversing all vertices, each \textit{buffer set} forms a candidate fusion plan.
\NAME will use the more accurate cost model \textit{latency-evaluator} to estimate 
which candidate plan performs the best.
The best one is the final fusion plan.

\subsection{Fusion Evaluation: Delta-Evaluator}
\label{subsec:fusion-eval}
\vspace{-.5ex}

\NAME applies \textit{delta-evaluator} to form the score function $f$.
One insight is that, we only need to estimate the performance gain or loss when forming a fusion pattern, 
but do not require accurate estimation of the overall execution time.
With this insight, the score function $f$ represents the performance gain or loss of 
a fusion pattern only.

There are three main factors in \textit{delta-evaluator}: reduced memory access latency,
reduced CPU-GPU context switch overhead, and performance penalty of kernel fusion.
The score function $f$ is the summary of these three parts, shown in equation
\ref{eq:cost-model-II}.

\begin{equation}
\begin{aligned}
\label{eq:cost-model-II}
\vspace{-1ex}
f=T_{reduced\_mem} + T_{reduced\_calls} - T_{penalty}
\vspace{-1ex}
\end{aligned}
\end{equation}

We estimate reduced memory access latency ($T_{reduced\_mem}$) with two factors. 
The first is the amount of memory traffics between operators to be fused, which
can be calculated according to the shape and type of input and output tensors.
And the second is the change of memory type to store the intermediate values
between operations. We build a regression model to predict the reduced memory 
access latency when changing the memory type from global memory to register
or shared memory, when given memory traffic amount. 
The regression model is based on latency data we collected offline 
on various GPU vendors with various amount of data traffic. 




$T_{reduced\_calls}$ can be easily estimated by multiplying the number of fused
kernels with a fixed value representing average CPU-GPU context switch time we collected.

The performance penalty ($T_{penalty}$) is estimated 
by a simplified version of \textit{latency-evaluator} (sec~\ref{subsec:cost-model-I}).
As fusion plan exploration faces a very large searching space, 
using \textit{latency-evaluator} directly takes too long time for JIT optimization.
The most time consuming part of \textit{latency-evaluator} is the estimation of $Occupancy$.
To estimate $Occupancy$ fast, we use a fixed number (16) as registers number, 
and use the maximal shared memory usage in and between any ops within a fusion pattern 
as the overall shared memory usage.
Life time analyzing of registers and shared memory is discarded in \textit{delta-evaluator}.

\vspace{-.5ex}
\section{Implementation}
\label{section:implementation}
\vspace{-1ex}

The fusion exploration and code generation techniques we studied requires 
heavy implementation efforts and will be a burden if left to users.
Meanwhile, TensorFlow evolves fast and porting the realization of \NAME 
from one TensorFlow version to another one is heavy.

To ease the using of \NAME and make it portable to any version of TensorFlow,
we build \NAME as a stand alone add-on.
Users can make use of all the techniques without changing any model script, 
but only need to specify \NAME path and setup environment to enable \NAME 
beyond TensorFlow XLA.

Specifically, we add \NAME as an extra pass upon XLA framework. 
After XLA finishes all its optimizations, including its basic fusion pass, 
we feed the fusion results of XLA into \textit{fusion explorer} to get larger scope fusions.
We do not disable the basic XLA fusion pass as 
there is no conflict to apply \NAME based on the naive fusions 
and a basic fusion reduces optimization time of \NAME.

We realize a GPU IR emitter module with the techniques of \textit{code generator} in \NAME.
The fusion plan generated by \textit{fusion explorer} goes through this IR emitter.
The IR emitter will compile fusion patterns into llvm IR, GPU ptx IR, 
and finally GPU binary sequentially.

Note that the input of \NAME is the basic fusion results of XLA, 
and not every basic fusion will be merged into a \NAME fusion pattern as a result.
The basic fusions not merged into larger fusions by \NAME will finally 
go through basic compilation pass of XLA.

All the above implementations are realized as a hook of TensorFlow,
and will be called if \NAME is loaded and environment is set.

To hide the optimization tuning time, 
we develop async-compilation mode in \NAME library.
As for training, if users set environment as async-optimization, 
the first several iterations will not execute binaries generated by \NAME 
while \NAME is still running in background asynchronously. 
After \NAME generates optimized kernels for the model, 
the non-optimized ops will be replaced with that of \NAME in latter iterations.



\vspace{-.5ex}
\section{Evaluation}
\label{section:evaluation}
\vspace{-1ex}

\subsection{Experimental Setup}
\label{subsec:eval-setup}
\vspace{-.5ex}

In this section, we evaluate \NAME using a variety of machine learning applications 
with different characteristics in different fields. Table \ref{tbl:workloads-for-evaluation} 
summarizes the various fields of the evaluated applications and the characteristics 
of these applications. These applications range from images (\emph{CRNN}\cite{shi2016end}), 
speech (\emph{ASR}\cite{yu2016automatic}), NLP (\emph{Transformer}\cite{vaswani2017attention},
\emph{BERT}\cite{devlin2018bert}), to internet scale E-commerce search and recommendation systems 
(\emph{DIEN}\cite{zhou2019deep}). The building blocks of these workloads include perceptron, 
attention, convolution, RNN and a broad range of memory intensive operators.

\begin{table}
\begingroup
\caption{Workloads for evaluation.}
\vspace{-3ex}
\label{tbl:workloads-for-evaluation}
\begin{center}
\begin{small}
\renewcommand{\arraystretch}{0.85}
\begin{tabular}{cccc}
\toprule
Model       & Field              & Mode      & Batch Size \\
\midrule
BERT        & NLP                & Both      & 32         \\
DIEN        & Recommendation     & Both      & 256 \\
Transformer & NLP                & Training  & 4096       \\
ASR         & Speech Recognition & Inference & 8          \\
CRNN        & OCR                & Inference & 8          \\
\bottomrule
\end{tabular}
\end{small}
\end{center}
\endgroup
\end{table}

To demonstrate the benefits of \NAME over previous work, we compare it with the default 
TensorFlow implementation and XLA (up-to-date with community functions).
All evaluation results are collected on NVIDIA V100 GPU with 16 GB device memory. The server
runs Red Hat Enterprise Linux 7.2 with CUDA toolkit 10.2 and cuDNN 7.6.

\subsection{Overview}
\label{subsec:end-to-end}
\vspace{-.5ex}

We evaluate the speedup of \NAME by comparing inference cost or the training time of one 
iteration for TF, XLA and \NAME with the same batch-size. 
During our test, the accuracy in each iteration of training and the result of 
inference are the same with TF and XLA.
We repeat 10 times and use the average performance to validate speedup. 
As for training workloads, we collect the execution time from the 11th iteration to the 20th (guaranteed to be stable), 
to avoid the initialization overhead of the early training iterations.

\begin{figure}
    \centering
    \includegraphics[width=\columnwidth]{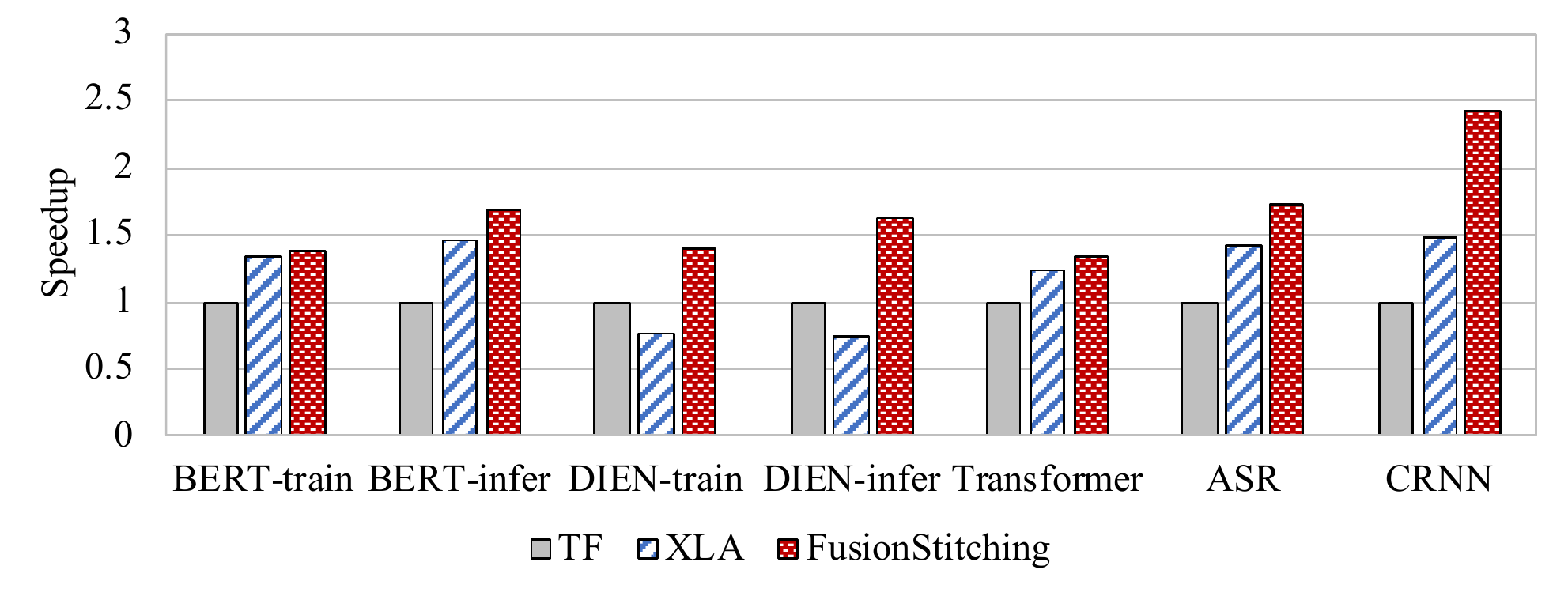}
    \vspace{-3ex}
    \caption{Performance speedup of \NAME.}
    \vspace{-2ex}
    \label{fig:overall-speedup}
\end{figure}

We show the speedup of \NAME in figure \ref{fig:overall-speedup}, 
where the execution time of TensorFlow is normalized to 1.
Compared to TensorFlow, our approach achieves up to $2.42\times$ speedup, 
with 1.66$\times$ on average. 
Compared to XLA, our approach achieves up to $2.21\times$ speedup, with 1.45$\times$ on average. 
Note XLA shows performance degradation for \emph{DIEN} (both training and inference), 
while \NAME does not show negative optimization in any of these cases.

We also test the inference workloads on NVIDIA T4 GPU and get the similar speedup.

We apply \NAME in production and measure the performance benefits. 
It shows that \NAME saves about 7,000 GPU hours for about 30,0000 tasks in a month.
The performance result on the cluster shows \NAME is robust.
Note that XLA, as state-of-the-art JIT engine, cannot be enabled default 
as it suffers from negative optimizations for many cases.

Overall, the performance benefit of \NAME comes from both reduced CPU-GPU context switch 
and off-chip memory access. 
\NAME gives a good solution to the problems we found (sec \ref{subsec:motivation}).
We provide performance breakdown information in the following section.

\subsection{Performance Breakdown}
\label{subsec:breakdown}
\vspace{-.5ex}

Table \ref{tbl:kernel-breakdown} shows the kernel breakdown information, 
including execution time (\textit{T}) of memory intensive ops (\textit{Mem}), 
compute intensive ops (\textit{Math}), CPU time (\textit{CPU}, kernel launch and framework scheduling), 
CUDA memcpy/memset activities (\textit{Cpy}) and kernel call times (\textit{\#}). 
Note that the breakdown profiling process is different with the process to measure the 
end-to-end performance. This is because profiling introduces some overhead and makes 
the end-to-end time not accurate.
We profile \textit{Mem}, \textit{Math}, \textit{Cpy} time directly, 
and get \textit{CPU} time by dividing other time from end-to-end time.

Before we analyze the effect of our technique, we need to point out that XLA 
affects the behavior of Matrix operations (GEMM and GEMV). It tends to fuse 
GEMVs into GEMM, and GEMMs to larger GEMM when there are Matrices sharing common input. 
Some other algebra transformation and loop-invariant code motion also reduces GEMM count. 
The difference in GEMM count in table \ref{tbl:kernel-breakdown} is caused by such reasons.
Meanwhile, XLA affects the runtime behavior of TensorFlow and leads to more or 
less CUDA memcpy/memset activities. \NAME is built upon 
XLA and exhibits the same behavior.

\begin{table}[!htb]
\caption{Kernel execution breakdown. 
\textit{TF}: naive TensorFlow. 
\textit{FS}: \NAME. 
\textit{CPU}: the scheduling and pre-/post-processing metrics on CPU. 
\textit{Math}: compute-intensive kernels. 
\textit{Mem}: memory-intensive kernels. 
\textit{Cpy}: CUDA memcpy and memset calls.
\textit{E2E}: the end-to-end time of one iteration (in milliseconds).
\textit{T}: execution time.
\textit{\#}: kernel calls number.}
\vspace{-3ex}
\label{tbl:kernel-breakdown}
\begin{center}
\begin{small}
\setlength{\tabcolsep}{3pt} 
\renewcommand{\arraystretch}{0.88} 
\begin{tabular}{lccccccc}
\toprule
Model                          & \multicolumn{1}{l}{Tech} & T/\# & CPU & Math & Mem  & Cpy &  E2E \\
\midrule
\multirowcell{6}{BERT\\-train} & \multirow{2}{*}{TF}  & T  & 1.55 & 41.69 & 28.45 & 0.15 & 71.84 \\
                               &                      & \# & - & 98 & 561 & 102  & 761 \\
                               & \multirow{2}{*}{XLA} & T  & 2.3 & 41.89 & 9.56 & 0.15 & 53.9 \\
                               &                      & \# & - & 98 & 200 & 97  & 395 \\
                               & \multirow{2}{*}{FS}  & T  & 2.8 & 42.11 & 7.02 & 0.03 & 51.96 \\
                               &                      & \# & - & 98 & 98 & 20  & 216 \\
\midrule
\multirowcell{6}{BERT\\-infer} & \multirow{2}{*}{TF}  & T  & 3.24 & 1.65 & 0.83 & 0.14 & 5.86 \\
                               &                      & \# & - & 70 & 365 & 106  & 541 \\
                               & \multirow{2}{*}{XLA} & T  & 0.78 & 2.50 & 0.60 & 0.13 & 4.02 \\
                               &                      & \# & - & 98 & 277 & 94  & 469 \\
                               & \multirow{2}{*}{FS}  & T  & 0.59 & 2.46 & 0.40 & 0.04 & 3.49 \\
                               &                      & \# & - & 98 & 77 & 30  & 205 \\
\midrule
\multirowcell{6}{DIEN\\-train} & \multirow{2}{*}{TF}  & T  & 90.13 & 7.77 & 32.54 & 7.12 & 137.56 \\
                               &                      & \# & - & 1218 & 10406 & 1391  & 13015 \\
                               & \multirow{2}{*}{XLA} & T  & 124.04 & 9.06 & 37.50 & 6.56 & 177.16 \\
                               &                      & \# & - & 1215 & 6842 & 1996  & 10053 \\
                               & \multirow{2}{*}{FS}  & T  & 48.42 & 7.91 & 35.84 & 5.55 & 97.72 \\
                               &                      & \# & - & 1215 & 2109 & 1395  & 4719 \\
\midrule
\multirowcell{6}{DIEN\\-infer} & \multirow{2}{*}{TF}  & T  & 27.36 & 2.58 & 7.55 & 1.99 & 39.48 \\
                               &                      & \# & - & 406 & 3680 & 225  & 4311 \\
                               & \multirow{2}{*}{XLA} & T  & 44.21 & 2.24 & 6.12 & 0.94 & 53.51 \\
                               &                      & \# & - & 405 & 2585 & 627  & 3617 \\
                               & \multirow{2}{*}{FS}  & T  & 17.54 & 2.45 & 3.51 & 0.7 & 24.20 \\
                               &                      & \# & - & 405 & 815 & 422  & 1642 \\
\midrule
\multirowcell{6}{Trans\\former} & \multirow{2}{*}{TF} & T  & 5.92 & 107.04 & 81.03 & 1.14 & 195.37 \\
                               &                      & \# & - & 399 & 2497 & 522 & 3418 \\
                               & \multirow{2}{*}{XLA} & T  & 11.59 & 106.58 & 38.33 & 1.20 & 157.70 \\
                               &                      & \# & - & 405 & 903 & 577 & 1885 \\
                               & \multirow{2}{*}{FS}  & T  & 7.43 & 107.16 & 30.26 & 0.80 & 145.65 \\
                               &                      & \# & - & 400 & 423 & 369 & 1212 \\
\midrule 
\multirowcell{6}{ASR}         & \multirow{2}{*}{TF}   & T  & 9.36 & 2.84 & 3.06 & 0.63 & 15.89  \\
                              &                       & \# & -     & 76 & 1359 & 439  & 1942  \\
                              & \multirow{2}{*}{XLA}  & T  & 6.61 & 0.09 & 4.00 & 0.40 & 11.10 \\
                              &                       & \# & -     & 4 & 386 & 257 & 647     \\
                              & \multirow{2}{*}{FS}   & T  & 5.51  & 0.11 & 3.10 & 0.45 & 9.18   \\
                              &                       & \# & -     & 4 & 187 & 284  & 475     \\
\midrule
\multirowcell{6}{CRNN}        & \multirow{2}{*}{TF}   & T  & 23.31 & 6.05 & 6.14  & 1.60 & 37.10 \\
                              &                       & \# & -     & 256  & 3674 & 890  & 4820  \\
                              & \multirow{2}{*}{XLA}  & T  & 12.17 & 0.30 & 11.37 & 1.04 & 24.88 \\
                              &                       & \# & -     & 7    & 993   & 406  & 1406  \\
                              & \multirow{2}{*}{FS}   & T  & 6.35  & 0.31 & 7.69  & 1.01 & 15.36 \\
                              &                       & \# & -     & 8    & 311   & 388  & 707   \\
\bottomrule
\end{tabular}
\end{small}
\end{center}
\end{table}

According to Table \ref{tbl:kernel-breakdown}, we have the following observations.

\parahead{Reduced context switch overhead.}
\NAME effectively reduces the memory intensive kernel calls of all workloads, which results in 
reduced kernel launch and framework scheduling overhead. As is shown in Table \ref{tbl:kernel-breakdown}, 
the number of memory intensive kernel calls with \NAME is 38.0\% of that with XLA in average,
ranging from 27.8\% to 48.4\%. 
Meanwhile, \NAME reduces CUDA memcpy/memset activities (\textit{Cpy} time) than XLA, with 34.3\% decrease in average.
This is because we fuse more ops together into larger kernels than XLA and 
many CUDA memcpy/memset are combined together.
The \textit{CPU} time difference in table \ref{tbl:kernel-breakdown} 
indicates the reduced time due to the decrease of kernel calls and CUDA memcpy/memset activities. 
\NAME achieves up to 61.0\% saving of the $CPU$ time comparing with XLA, 41.0\% in average. 

Take \textit{DIEN-train} as an example, the kernel call number for memory intensive ops 
is 2109 with \NAME, and 6842 with XLA. Meanwhile, the CUDA memcpy/memset activities 
is reduced to 1395, comparing to 1996 with XLA.
The final $CPU$ time with \NAME is significantly less than both TF and XLA, 
thanks for the reduced kernel calls.
Note that XLA increases CUDA memcpy/memset activities and results in severe performance drop here. 
\NAME avoids the increased memcpy/memset calls due to larger kernel granularity 
and do not suffer from the drawback. 
\textit{DIEN-infer} has the similar behavior.
Optimizing kernel fusions while considering runtime behaviors (like memcpy activities) could be a future 
research topic.

\parahead{Reduced memory intensive op execution time.}
\NAME reduces the total execution time for memory-intensive operations. 
The speedup of memory-intensive ops for all workloads is $1.39\times$ in average comparing with XLA, 
and up to $1.74\times$. 
The performance speedup mainly comes from reduced global memory access. 
By fusing memory-intensive operations aggressively, the intermediate values 
can be cached in registers and shared memory.

Take CRNN as an example, it reads 667.6 MB global memory with XLA, 
while \NAME reduces the traffic to 225.8 MB. 
About 66\% global memory access has been reduced for memory intensive ops.
The execution time of all memory intensive computations thus achieves 
$1.48\times$ speedup than XLA.


Overall speaking, \NAME supports more complex fusion patterns than XLA with 
effective kernel generation, which relaxes the fusion conditions and thus reduces 
the final kernel numbers and intermediate global memory transactions. 
These two factors are essential for the performance of memory intensive ops (sec~\ref{subsec:motivation}).
Meanwhile, with well controlled performance estimation and reduced runtime memcpy activities than XLA, 
\NAME is less likely to result in bad case about optimization.

\subsection{Fusion Pattern Analysis}
\vspace{-.5ex}

We take the \textit{Layer Normalization} case in Figure~\ref{fig:fusion-pattern-case} again 
to analyze the fusion result of XLA and \NAME.
Note \textit{Layer Normalization} is a very common component in deep learning models.

XLA forms four different fusion kernels. 
While \NAME is able to fuse them all and generate a more efficient kernel.
We collect the performance of all kernels of the two version. 
The single kernel with \NAME achieves a speedup of $1.23\times$ 
comparing with the sum of all 4 kernels with XLA.
For this test, we do not count the context switch overhead for the 4 kernels in XLA version, 
which further hurts the performance of XLA.

There are two factors that prevents 
XLA to fuse operators with a larger granularity in this case. 
The first reason is \textit{reduce} op in \textit{xla-fusion.7} and \textit{xla-fusion.3}.
As we discussed before, XLA does not explore reuse of intermediate results of producer op.
Fusing reduce op in the middle of a kernel results redundant computation 
of the same value in different threads, thus hurts the performance.
The second reason is expensive element-wise ops with small tensor shape (\textit{xla-fusion.2}). 
XLA does not tend to fuse expensive ops processing small tensors in the middle of a kernel 
as the fusion will introduce redundant computations of expensive instructions.

Instead, \NAME finds notable potential for larger granularity fusion 
in this case and fuses all operators into one kernel. 
It applies data reuse to prevent redundant computation.
In this way, the intermediate global memory transaction and CPU-GPU context switch is avoided.

\subsection{Overhead Analysis}
\vspace{-.5ex}

Similar with XLA, \NAME is designed for tune-once-run-many-times scenarios, 
which is a basic characteristic of deep learning workloads.
For the training that could take up to several days, \NAME 
only needs to run in the first training iteration.
For an inference task, the executable kernels can be prepared once 
and run many times in the future.

We measure the one-time overhead introduced by \NAME compared to XLA for training.
The value is the JIT compilation time of \NAME minus that with XLA. 
Results show that the extra overhead is less than 30 minutes for the workloads 
we evaluated in this paper, which is far less than the overall training time.

As for cost models, we tried to use complete \textit{latency-evaluator} 
to estimate $T_{penalty}$ in \textit{delta-evaluator}. 
The results show a much longer tuning time, but do not show better performance of tuning results.
It demonstrates the simplified cost model works good for fusion exploration.

A problem of both \NAME and XLA is that, they cannot handle dynamic shapes, 
appears in some deep learning workloads, with low tuning overhead. 
The reason is that the design of XLA service framework is not friendly to dynamic shape, 
while \NAME is implemented based on XLA service framework. 
This implementation problem does not affect the insight that \NAME shows.




\vspace{-.5ex}
\section{Related Work}
\label{section:relatedwork}
\vspace{-1ex}

Many current AI compilation optimization works mainly focus on 
compute intensive ops\cite{tvm,halide,ma2020rammer,boda,cowan2020automatic,kim2019code,latte},
but do not pay attention to memory intensive ops.
XLA\cite{xla} is one work optimizing memory intensive ops in depth. 
It provides an just-in-time fusion engine to reduce context switch 
and off-chip memory access overhead. 
However, XLA lacks to explore data reuse of intermediate value and 
limits the fusion exploration space.
Besides, it uses simple rules to explore fusion strategy, 
which is not adaptive to varied combinations of ops and tensor shapes.
\NAME reveals the data reuse opportunity for JIT fusion and 
proposes cost-based approach for fusion searching and code generation tuning.

Some works study about fusion optimization of machine learning workloads for static patterns.
Li et al.\cite{li2020automatic} explores horizontal fusion for GPU Kernels to 
increase thread-level parallelism.
Appleyard et al.\cite{appleyard2016optimizing}, Diamos et al.\cite{diamos2016persistent} study 
about kernel fusion technique to speedup RNN workloads. 
These works do not study about just-in-time fusion problem that processes 
fusion patterns unknown ahead. 
Neither do they explore op combination strategy.
Abdolrashidi et al.\cite{abdolrashidi2019learning} explores op combination strategies 
with Proximal Policy Optimization\cite{schulman2017proximal} algorithm. 
This is an ahead-of-time tuning approach that searches in 
simple fusion rules similar with XLA, and does not explore fusion of complex patterns.

Some works focusing on compute intensive ops have ability of op fusion.
Recent TVM\cite{tvm} and Ansor\cite{zheng2020ansor} implementation 
have fusion ability with simple rules for simple patterns. 
These two works usually requires a long time tuning process 
given a static computation description, 
but are not designed to handle varied combinations of ops timely. 
Tiramisu\cite{baghdaditiramisu} applies a similar fusion approach with TVM.
Tensor Comprehensions\cite{tc} provides a polyhedral based JIT compiler capable of fusing ops together.
It focuses on the trade-off of parallelism and data locality.
Glow\cite{rotem2018glow} and Latte\cite{latte} supports basic fusion 
and does not explore complex combination scenarios.
Astra\cite{sivathanu2019astra} and Ashari et al.'s work\cite{ppopp2015} support GEMM and basic element-wise op fusion. 
None of these works reveal the fusion chance with intermediate data reuse 
between to-be-fused ops through shared memory and register shuffle just-in-time. 
They also lacks of a thoroughly study about just-in-time fusion strategy exploration.

Fusion approach has also been applied to a wide range of domains 
given a static computation description, 
like HPC\cite{hpcfusion}, database\cite{kernelweaver}, image processing\cite{cgo2019,scopes}.
They meet different challenges comparing with just-in-time compilation of AI workloads.
Specifically, they do not fuse kernels that are not known ahead 
and do not face the choice of fusing which ops together in a huge searching space.

CUDA Graph\cite{gray2019getting} reduces kernel launch overhead but suffers from 
severer initialization overhead and large extra GPU memory usage 
due to graph creation\cite{parravicini2020dag}.
It does not reduce off-chip memory traffic.

Performance models for GPU and other accelerators is another related research topic 
\cite{kaufman2020learned,lym2019delta,yang2019performance,
cui2012accurate,zhang2011quantitative,zhang2017understanding}.
Yet we design a domain specific cost model system for fusion 
and code generation requirements.



\vspace{-.5ex}
\section{Conclusion}
\label{section:conclusion}
\vspace{-1ex}

This work tackles the problem of optimizing memory intensive operators 
in machine learning workloads. 
We show that memory intensive ops are vital 
to end-to-end performance of various deep learning models. 
We propose \NAME that supports to fuse operators, 
with complex dependence and non-homogeneous parallelism, 
to reduce memory access and context switch overhead just-in-time. 
\NAME broadens the fusion possibility beyond state-of-the-art JIT techniques 
by introducing reuse scheme.
\NAME consists of \textit{fusion explorer} and \textit{code generator}. 
The \textit{fusion explorer} selects candidate fusion patterns 
from the large searching space with appropriate computing complexity,
and produces a fusion plan with promising performance expected. 
We provide a set of kernel composition schemes, 
with which \textit{code generator} stitches operators and reuses intermediate values
with on-chip memory as much as possible,
and tunes the schedules to emit high performance GPU code for a given fusion pattern. 
A two-layer cost model helps the searching and tuning process of \NAME. 
Results show that \NAME outperforms state-of-the-art JIT techniques 
with up to 2.21$\times$ speedup, 1.45$\times$ on average.

\bibliographystyle{plain}
\bibliography{refs}

\end{document}